\documentclass[10pt,twocolumn]{IEEEtran}
\usepackage{verbatim}
\usepackage{amssymb}  
\usepackage{color}
\usepackage{bm}
\usepackage{xcolor}
\usepackage{algorithm}
\usepackage{algorithmic}
\usepackage{theorem}  
\newcommand{\tabincell}[2]{\begin{tabular}{@{}#1@{}}#2\end{tabular}}

\usepackage{cite}
\ifCLASSINFOpdf
  \usepackage[pdftex]{graphicx}
\else
  \usepackage[dvips]{graphicx}
\fi
\usepackage{amsmath}

\usepackage{algorithmic}

\usepackage{array}

\ifCLASSOPTIONcompsoc
  \usepackage[caption=false,font=normalsize,labelfont=sf,textfont=sf]{subfig}
\else
  \usepackage[caption=false,font=footnotesize]{subfig}
\fi

\usepackage{stfloats}

\usepackage{url}

\theoremstyle{plain}

\theoremstyle{plain}
\newtheorem{theorem}{Theorem}
\theoremstyle{plain}

\theoremstyle{plain}

\theoremstyle{plain}

\theoremstyle{plain}

\begin{document}

% Put a box at the end of the proof
\def\QEDclosed{\mbox{\rule[0pt]{1.3ex}{1.3ex}}}
\def\QEDopen{{\setlength{\fboxsep}{0pt}\setlength{\fboxrule}{0.2pt}\fbox{\rule[0pt]{0pt}{1.3ex}\rule[0pt]{1.3ex}{0pt}}}}
\def\QED{\QEDopen}
\def\proof{}
\def\endproof{\hspace*{\fill}~\QED\par\endtrivlist\unskip}

\title{Partial Reciprocity-based Precoding Matrix Prediction in FDD Massive MIMO with Mobility}

\author{Ziao~Qin, Haifan~Yin,~\IEEEmembership{Senior Member,~IEEE} 
        %John~Doe,~\IEEEmembership{Fellow,~OSA,}
       % and~Jane~Doe,~\IEEEmembership{Life~Fellow,~IEEE}% <-this % stops a space
\thanks{Z. Qin, and H. Yin are with School of Electronic Information and Communications, Huazhong University of Science and Technology
430074 Wuhan, China (e-mail: ziao\_qin@hust.edu.cn, yin@hust.edu.cn.)}

\thanks{This work was supported by the National Natural Science Foundation of China under Grant 62071191. The corresponding author is Haifan Yin.}}

\maketitle

\begin{abstract}
The timely precoding of frequency division duplex (FDD) massive multiple-input multiple-output (MIMO) systems is a substantial challenge in practice, especially in mobile environments. In order to improve the precoding performance and reduce the precoding complexity, we propose a partial reciprocity-based precoding matrix prediction scheme and further reduce its complexity by exploiting the channel gram matrix. We prove that the precoders can be predicted through a closed-form eigenvector interpolation which was based on the periodic eigenvector samples. Numerical results validate the performance improvements of our schemes over the conventional schemes from 30 km/h to 500 km/h of moving speed.
\end{abstract}

% Note that keywords are not normally used for peer review papers.
% \fi
\IEEEpeerreviewmaketitle
\section{Introduction}
\IEEEPARstart{I}{n} a practical massive multiple-in multiple-out (MIMO) system with multiple antennas at the user equipment (UE), the downlink (DL) precoding inevitability introduces overwhelming high-dimensional singular value decomposition (SVD) operations. Considering the limited computation capability at the base station (BS), the timely precoding is highly challenging in mobile environments. This problem can be more serious in frequency division duplex (FDD) massive MIMO systems due to the asymmetry frequency bands, pilot training overhead and the limited feedback resources. %The non-unique nature of the SVD further challenges the precoding matrix prediction. 

The state-of-the-art schemes usually first estimated the DL channel matrix and then compute the precoding matrix directly from it in every subframe, called ``full-time SVD'' precoding scheme \cite{2022ziao,2022Fan}. These methods were found to be challenging to timely update the precoders in practical systems with limited computation ability.
In current communication systems \cite{3gpp214}, the precoding matrix was often updated in a periodic way to reduce the precoding complexity. However, the system spectral efficiency was observed to decline significantly in mobile environments. The alternative method is to leverage the precoder interpolation scheme to reduce the precoding complexity. The authors in \cite{2020gramesti} addressed the eigenvector interpolation in time frequency division (TDD) massive MIMO to reduce complexity based on the channel correlation across subcarriers. The authors in \cite{2021Flag} introduced a flag manifold-based precoder interpolation in FDD to reduce the feedback overhead. Nevertheless, the researchers in \cite{2020gramesti,2021Flag} did not consider the essential influence of mobility on the system. 

In order to reduce the precoding complexity and achieve a timely precoding in FDD systems, we exploit the partial reciprocity of the wideband channel matrix or the channel gram matrix and predict the precoding matrix through an exponential model consisting of several periodic channel eigenvector samples and the channel prediction results. The proposed scheme in this paper is based on the periodic SVD precoding scheme and differs from our prior full-time SVD precoding scheme in FDD \cite{2022ziao}. Our prior work proposed an eigenvector prediction (EGVP) method to interpolate the precoding matrix in TDD mode with the closed-form channel weight interpolation and channel prediction. Unfortunately, the proposed framework in \cite{qin2023eigenvector} can not work in FDD mode owing to the limited feedback channel state information (CSI) and the asymmetry frequency bands. To the best of our knowledge, few studies have investigated the precoding matrix prediction through an exponential model in FDD massive MIMO systems with mobility.
The main contributions are 
\begin{itemize}
    \item We propose a partial reciprocity-based wideband channel matrix eigenvector prediction (EGVP-PRWCM) scheme and further reduce its complexity by another scheme called partial reciprocity-based channel gram matrix eigenvector prediction (EGVP-PRCGM) with a slight performance loss caused by the channel estimation error.
    \item We prove that the precoding matrix can be predicted by a closed-form eigenvector interpolation based on the partial reciprocity in FDD massive MIMO. The upper bound of the precoding matrix prediction error is derived, which is related to the UL channel estimation power loss.
    \item We prove that the precoding complexity of the proposed schemes is reduced compared to the traditional full-time SVD scheme. Moreover, the numerical results demonstrate that the SE of the two schemes approach that of the full-time SVD scheme in mobile scenarios with speeds ranging from 30 km/h to 500 km/h.
\end{itemize}

\section{DL Channel estimation}\label{sec1}
In commercial massive MIMO systems with multiple antennas at the UE side, eigen zero-forcing (EZF) is a common method used in industry to acquire the DL precoders \cite{2010TSPeigenvalue}. The basic idea of EZF is to find the precoding matrix for the UEs through the SVD or eigenvalue decomposition (EVD) of the DL channel estimation results. Therefore, the DL channel estimation procedure is introduced first.

We consider a wideband FDD massive MIMO system. 
%equipped with multiple antennas at both the BS and the UE sides. 
The BS performs periodic SVD of the CSI once every $T_{\rm{svd}}$ ms to obtain the precoding matrix with EZF. The number of uniform planar array (UPA) antennas at the BS is $N_t$ and the number of UE antennas is $N_r$. All $K$ moving  UEs are randomly distributed within the cell. The superscripts $u,d$ denote the uplink (UL) and DL parameters, respectively. The wideband system has $N_f$ subcarriers. ${\Delta _t}$ denotes the length of one subframe. The subscript $r$ denotes the antenna at the UE. To simplify the presentation, the subscript $k$ is omitted in the following.
According to the channel model used in industry \cite{3gpp104}, the DL wideband channel with $P$ paths is
\begin{small}
    \begin{equation}\label{channel definition}
        {{\bf{h}}^d_r}\left( t \right) = \sum\limits_{p = 1}^P {\beta _{r,p}^d{e^{ - j2\pi f_0^d\tau _{r,p}^d}}{e^{jw_{r,p}^dt}}{\bf{r}}_{r,p}^d} ,
    \end{equation}
\end{small}where ${{\bf{r}}_p^u}$ is the angle-delay structure obtained by the Kronecker product of the delay and steering vector. $\beta _{r,p}$ is the complex amplitude. $f_0^d$ is the center frequency and $\tau _{r,p}^d$ is the delay parameter. The angular frequency ${w_{r,p}^d}$ denotes the Doppler frequency shift. 
Many state-of-the-art DL channel estimation frameworks can be applied, such as joint spatial division multiplexing (JSDM) \cite{2013JSDM}, compressed sensing \cite{2019SBL}, deep learning \cite{2021deep}, and joint-angle-delay-Doppler (JADD) \cite{2022ziao}. Due to its distinct performance in high mobility scenarios, we apply the JADD method in \cite{2022ziao} to obtain the DL CSI. During the estimation, $N_s$ angle-delay vectors are selected and the corresponding index set ${{\cal S}}_r^u$ is chosen by

\begin{small}
  \begin{equation}\label{angle-delay index selection}
    {\mathcal{S}}_r^u = \mathop {\arg \min }\limits_{\left| {{\mathcal{S}}_r^u} \right|} \left\{ {{\mathbb{E}}\left\{ {\left\| {{{\tilde {\mathbf{h}}}^u}\left( t \right)} \right\|_2^2} \right\} \geqslant \eta {\mathbb{E}}\left\{ {\left\| {{{\mathbf{h}}^u}\left( t \right)} \right\|_2^2} \right\}} \right\}.
\end{equation}   
\end{small}The UL channel estimation ${{{\tilde {\mathbf{h}}}^u}\left( t \right)}$ is achieved by selecting $N_s$  from $N_tN_f$ angle-delay vectors in the UL channel ${{{\mathbf{h}}^u}\left( t \right)}$. The threshold $\eta$ denotes the UL channel estimation power loss. 

Based on the partial reciprocity of the channel,
the DL channel with $T_d$ CSI delay is predicted at the BS:
\begin{small}
 \begin{equation}\label{DL prediction}
\tilde {\mathbf{h}}_r^d\left( {t + {T_d}} \right) = \sum\limits_{n \in {\mathcal{S}}_r^d} {\sum\limits_{m = 1}^M {\tilde a_{r,m}^d\left( n \right){e^{jw_{r,m}^d\left( n \right)\left( {t + {T_d}} \right)}}{{\mathbf{q}}_n}} },
\end{equation}   
\end{small}where ${{{\mathbf{q}}_n}}$ is the $n$-th column vector of a $N_tN_f$ sized DFT matrix $\bf{Q}$ and the corresponding index set is ${\mathcal{S}}_r^d$. Each ${{{\mathbf{q}}_n}}$ corresponds to $M$ DL Doppler $e^{jw_{r,m}^d\left( n \right)}$. ${\tilde a_{r,m}^d\left( n \right)}$ is the feedback amplitude. Moreover, $\kappa  = {{{N_s}M} \mathord{\left/
 {\vphantom {{{N_s}M} {{N_f}{N_t}}}} \right.
 \kern-\nulldelimiterspace} {{N_f}{N_t}}}$ is the ratio of the reduced dimension to the full dimension of the feedback.

%Once the DL CSI is obtained, our precoding matrix prediction schemes are shown in the following section.
\section{Precoding matrix prediction scheme}\label{sec2}
%In FDD mode, the DL channel estimation is achieved by a compressed feedback \eqref{DL prediction} and the same method in \cite{qin2023eigenvector} may not be applicable. 
In this section, we prove that the eigenvector can be interpolated by an exponential model and propose two precoding matrix prediction schemes based on the partial reciprocity in FDD mode. The general idea lies in decomposing the precoders into the channel weights and channels. 

The precoders obtained from the SVD of the wideband channel matrix ${\tilde{\bf{H}}^d}\left( {t} \right)$ equal to the eigenvectors of ${{\cal H}}\left( t \right) = {\tilde{\bf{H}}^d}\left( t \right){\tilde{\bf{H}}^d}{\left( t \right)^H}$ 
\begin{small}
    \begin{equation}\label{evd definition}
    {{\cal H}}\left( {t} \right){\tilde{\bf{u}}}_r^d\left( {t} \right) = {\chi _r}\left( {t} \right){\tilde{\bf{u}}}_r^d\left( {t} \right),\bmod \left( {t - {t_{{\text{in}}}},{T_{{\text{svd}}}}} \right) = 0,
\end{equation}
\end{small}where ${\tilde{\bf{H}}^d}\left( {t} \right) = \left[ {\begin{array}{*{20}{c}}
{\tilde{\bf{h}}_1^d\left( {t} \right)}&{\tilde{\bf{h}}_2^d\left( {t} \right)}& \cdots &{\tilde{\bf{h}}_{{N_r}}^d\left( {t} \right)}
\end{array}} \right]$ and ${\chi _r}\left( {t} \right)$ is the $r$-th eigenvalue. The initial subframe is $t_{\rm{in}}$. The corresponding eigenvector ${\tilde{\bf{u}}}_r^d\left( {t} \right)$ is decomposed into a linear combination of the channel weights and channels 
\begin{small}
  \begin{equation}\label{weight sample}
     {\tilde{\bf{u}}}_r^d\left( t \right) = \sum\limits_{j = 1}^{{N_r}} { \lambda _{r,j}^d\left( t \right) \tilde{\bf{h}}_j^d\left( t \right)} ,
\end{equation}  
\end{small}where the channel weight $\lambda _{r,j}^d\left( t \right)$ can be estimated with an exponential model according to Theorem 1 in \cite{qin2023eigenvector}, i.e., $\lambda _{r,j}^d\left( t \right) = \sum\limits_{l = 1}^{{L_{r,j}}} {b_{r,j}^d\left( l \right){e^{jw_{r,j}^d\left( l \right)t}}}$. The eigenvectors is then calibrated to remove the EVD uncertainty
\begin{small}
   \begin{equation}\label{eigenvector phase align}
    \left[ {\begin{array}{*{20}{c}}
  {\tilde {\mathbf{u}} _r^d\left( {{t_{{\text{in}}}}} \right)} \\ 
  {\tilde {\mathbf{u}} _r^d\left( {{t_{{\text{in}}}} + {\Delta _t}} \right)} \\ 
   \cdots  \\ 
  {\tilde {\mathbf{u}} _r^d\left( {{t_{{\text{ed}}}}} \right)} 
\end{array}} \right] = \left[ {\begin{array}{*{20}{c}}
  {\tilde {\mathbf{u}} _r^d\left( {{t_{{\text{in}}}}} \right)} \\ 
  {{\Delta _r}\left( {{t_{{\text{in}}}} + {\Delta _t}} \right)\tilde {\mathbf{u}} _r^d\left( {{t_{{\text{in}}}} + {\Delta _t}} \right)} \\ 
   \cdots  \\ 
  {{\Delta _r}\left( {{t_{{\text{ed}}}}} \right)\tilde {\mathbf{u}} _r^d\left( {{t_{{\text{ed}}}}} \right)} 
\end{array}} \right],
\end{equation} 
\end{small}where $t_{\rm{ed}}$ denotes the end subframe and the phase shift is ${\Delta _r}\left( t \right) = {\tilde {\mathbf{u}} _r}{\left( t \right)^H}{\tilde {\mathbf{u}} _r}\left( {{t_{{\text{in}}}}} \right)$. 
 
\begin{figure}[!t]
\centering
\includegraphics[width=2.8in]{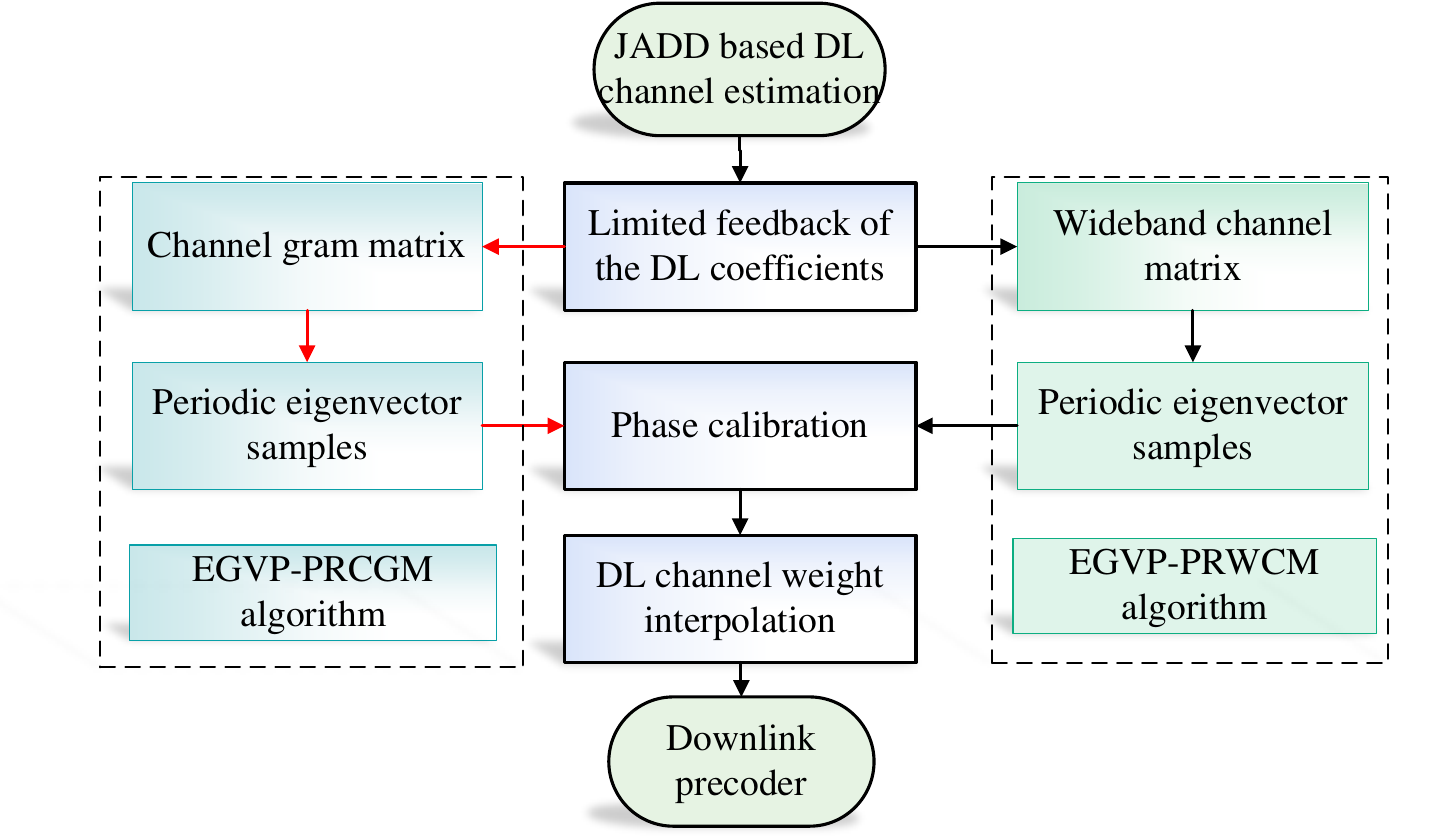}
\caption{The framework of the proposed partial reciprocity-based precoding matrix prediction schemes.}
\vspace{-0.3cm}
\label{flowchart}
\end{figure}
Fig. \ref{flowchart} demonstrates the framework of two proposed precoding matrix prediction schemes. First, the DL channel is estimated by the JADD method. Based on the limited feedback of the DL channel coefficients, the EGVP-PRWCM utilizes the wideband channel matrix to obtain eigenvector samples, whereas the EGVP-PRCGM relies on the channel gram matrix. Once the eigenvector samples have been obtained and calibrated, the channel weight is interpolated. Finally, the DL precoding matrix is predicted by \eqref{weight sample}.
\subsection{EGVP-PRWCM prediction scheme}
Rewrite Eq. \eqref{weight sample} in a matrix form
\begin{small}
 \begin{equation}\label{linear decompose of eigenvector}
    {{\tilde{\bf{U}}}^d}\left( {t} \right) = {\tilde{\bf{H}}^d}\left( {t} \right){\bf{\Lambda }}^d\left( {t} \right),
\end{equation}   
\end{small}where ${{\tilde{\bf{U}}}^d}\left( {t} \right) = \left[ {\begin{array}{*{20}{c}}
{{\tilde{\bf{u}}}_1^d\left( {t} \right)}&{{\tilde{\bf{u}}}_2^d\left( {t} \right)}& \cdots &{{\tilde{\bf{u}}}_{{N_r}}^d\left( {t} \right)}
\end{array}} \right]$ and ${\bf{\Lambda }}^d\left( {t} \right)$ is the channel weight matrix 
\begin{small}
 \begin{equation}\label{eigenvalue matrix}
    {\bf{\Lambda }}^d\left( {t} \right) = \left[ {\begin{array}{*{20}{c}}
{{\lambda _{1,1}^d}\left( {t} \right)}&{{\lambda _{2,1}^d}\left( {t} \right)}& \cdots &{{\lambda _{{N_r},1}^d}\left( {t} \right)}\\
{{\lambda _{1,2}^d}\left( {t} \right)}&{{\lambda _{2,2}^d}\left( {t} \right)}& \cdots &{{\lambda _{{N_r},2}^d}\left( {t} \right)}\\
 \cdots & \cdots & \cdots & \cdots \\
{{\lambda _{1,{N_r}}^d}\left( {t} \right)}&{{\lambda _{2,{N_r}}^d}\left( {t} \right)}& \cdots &{{\lambda _{{N_r},{N_r}}^d}\left( {t} \right)}
\end{array}} \right],
\end{equation}   
\end{small}where each column of ${\bf{\Lambda }}^d\left( {t} \right)$ is ${{\boldsymbol{\lambda}} _{r}^d\left( t \right)}$. In FDD mode, the DL channel is estimated by the compressed feedback of the CSI. The following theorem proves that the DL eigenvector can be asymptotically estimated by an exponential model.

\begin{theorem}\label{Theorem 1}
      When ${{N_t},{N_f} \to \infty }$, the eigenvectors obtained from the estimated DL channel in \eqref{evd definition} can be estimated by the following model.
      \begin{equation}\label{theorem Ns equation}
         \mathop {\lim }\limits_{{N_t},{N_f} \to \infty }  {\tilde{\bf{u}}}_r^d\left( t \right) = \sum\limits_{j = 1}^{{N_r}} {\sum\limits_{l = 1}^{{L_{r,j}}} {b_{r,j}^d\left( l \right){e^{jw_{r,j}^d\left( l \right)t}}} \tilde {\bf{h}}_j^d\left( t \right)} ,
      \end{equation}
      where the normalized mean square prediction error within $N_L$ subframes satisfies $\mathbb{E}{\left\{ {\frac{{\left\| {{{\mathbf{u}}_r^d\left( t \right)}-\tilde {\bf{u}}_r^d\left( t \right)} \right\|_2^2}}{{\left\| {{\bf{u}}_r^d\left( t \right)} \right\|_2^2}}} \right\}_{{N_L}}} \le 1- \eta $. The coefficients ${L_{r,j}}$, ${b_{r,j}^d\left( l \right)}$ and ${{e^{jw_{r,j}^d\left( l \right)}}}$ are the number of exponentials, amplitude and exponetials of the channel weight $\lambda _{r,j}^d\left( t \right)$, respectively.
\end{theorem}
\begin{proof}
\quad \emph{Proof:} Please refer to Appendix \ref{Appendix Theorem 1}.
\end{proof}

Theorem \ref{Theorem 1} gives a closed-form estimation model of the DL eigenvectors and a upper bound of the estimation error $1-\eta$, where $\eta$ is the UL channel estimation power loss. Therefore, in EGVP-PRWCM scheme, the precoding matrix prediction is transformed into the prediction of the channel weight which can be interpolated by $N_{\rm{svd}} \ge 2L_{r,j}$ eigenvector samples
\begin{small}
  \begin{equation}\label{channel weight interpolation}
    \hat \lambda _{r,j}^d\left( {{t_p}} \right) = \sum\limits_{l = 1}^{{L_{r,j}}} {b_{r,j}^d\left( l \right){e^{j\frac{{ w_{r,j}^d\left( l \right)}}{{{T_{{\text{svd}}}}}}{t_p}}}} ,
\end{equation}  
\end{small}where the interpolated subframe $t_p$ satisfies 
\begin{small}
  \begin{equation}\label{interpolated subframe}
    \bmod \,\left( {{t_p} - {t_{{\text{in}}}},{T_{{\text{svd}}}}} \right) \ne 0,{t_p} \in \left[ {{t_{{\text{in}}}},{t_{{\text{ed}}}} } \right].
\end{equation}  
\end{small}
At last, the predicted eigenvector is reconstructed
\begin{small}
   \begin{equation}\label{predicted eigenvector}
    \hat {\bf{u}}_r^d\left( t \right) = \sum\limits_{j = 1}^{{N_r}} {\hat \lambda _{r,j}^d\left( t \right)\tilde {\bf{h}}_j^d\left( t \right)}. 
\end{equation} 
\end{small}

In general, EGVP-PRWCM scheme relies on the periodic eigenvector samples obtained from the wideband channel matrix. However, the complexity order of EVD in this scheme is ${\cal{O}}\left( {{{\left( {{N_f}{N_t}} \right)}^3}} \right)$. It can be further reduced in the following EGVP-PRCGM scheme.
\subsection{EGVP-PRCGM prediction scheme}
The key to the complexity reduction of the EGVP-PRCGM scheme lies in calculating the channel weight samples by the EVD of the channel gram matrix $\tilde {\bf{S}}^d\left( {t} \right)$ instead of the wideband channel matrix ${{\cal H}}\left( t \right)$. The channel gram matrix $\tilde {\bf{S}}^d\left( {t} \right)$ is defined by
\begin{small}
     \begin{equation}\label{channel gram matrix definition}
    \tilde {\bf{S}}^d\left( {t} \right) = \left[ {\begin{array}{*{20}{c}}
{{{\tilde s}_{1,1}^d}\left( {t} \right)}&{{{\tilde s}_{2,1}^d}\left( {t} \right)}& \cdots &{{{\tilde s}_{{N_r},1}^d}\left( {t} \right)}\\
{{{\tilde s}_{1,2}^d}\left( {t} \right)}&{{{\tilde s}_{2,2}^d}\left( {t} \right)}& \cdots &{{{\tilde s}_{{N_r},2}^2}\left( {t} \right)}\\
 \cdots & \cdots & \cdots & \cdots \\
{{{\tilde s}_{1,{N_r}}^d}\left( {t} \right)}&{{{\tilde s}_{2,{N_r}}^d}\left( {t} \right)}& \cdots &{{{\tilde s}_{{N_r},{N_r}}^d}\left( {t} \right)}
\end{array}} \right].
\end{equation}   
\end{small}First, each element $\tilde s^d_{r,j}\left( t \right) = \left\langle {\tilde {\bf{h}}_r^d\left( t \right),\tilde {\bf{h}}_j^d\left( t \right)} \right\rangle$ can be equally written as
\begin{small}
\begin{align}\label{channel relevance reciprocity}
  {\tilde s_{r,j}^d\left( t \right)}&{ = \sum\limits_{n \in {{\mathcal{G}}_{r,j}}\left( t \right)}^{} {{{\left( {\sum\limits_{m = 1}^M {a_{j,m}^d\left( n \right){e^{jw_{r,m}^d\left( n \right)t}}} } \right)}^H}} {\text{ }}} \nonumber\\ 
  {}&{\left( {\sum\limits_{m = 1}^M {a_{r,m}^d\left( n \right){e^{jw_{r,m}^d\left( n \right)t}}} } \right),} 
\end{align}
\end{small}
where the set ${{\mathcal{G}}_{r,j}}\left( t \right) = {\mathcal{S}}_j^d\left( t \right) \cap {\mathcal{S}}_r^d\left( t \right)$ is the index intersection of the angle-delay vector of two channels.
Based on the channel partial reciprocity in FDD, the DL Doppler ${e^{jw_{r,m}^d\left( n \right)}}$, the amplitude ${a_{r,m}^d\left( n \right)}$ and the angle-delay vector index set ${\mathcal{S}}_r^d\left( t \right)$ are obtained by the UL channel parameter estimation utilizing the JADD method \cite{2022ziao}. 

Second, apply an EVD of $\tilde {\bf{S}}^d\left( {t} \right)$ and obtain the eigenvectors %${\tilde {\mathbf{\Lambda }}^d}\left( t \right) = \left[ {\begin{array}{*{20}{c}}{\tilde {\boldsymbol{\lambda }}_1^d\left( t \right)}&{\tilde {\boldsymbol{\lambda }}_2^d\left( t \right)}& \cdots &{\tilde {\boldsymbol{\lambda }}_{{N_r}}^d\left( t \right)} \end{array}} \right]$ 
%as the mapping results of the channel weight ${{\boldsymbol{\lambda}} _{r}^d\left( t \right)}$ in JADD
\begin{small}
 \begin{equation}\label{evd of channel gram matrix}
   \tilde {\bf{S}}^d\left( t \right){\tilde {\boldsymbol{\lambda }}_r^d}\left( t \right) = {\tilde \chi _r}\left( t \right){\tilde {\boldsymbol{\lambda }}_r^d}\left( t \right).
\end{equation}   
\end{small}Similar to the EGVP-PRWCM scheme, the eigenvector samples of the EGVP-PRCGM scheme are 
\begin{small}
  \begin{equation}\label{eigenvector sample from channel relevance}
    \tilde {\mathbf{u}}_{r,g}^d\left( t \right) = \sum\limits_{j = 1}^{{N_r}} {\tilde \lambda _{r,j}^d\left( t \right)\tilde {\mathbf{h}}_j^d\left( t \right)} ,\bmod \left( {t - {t_{{\text{in}}}},{T_{{\text{svd}}}}} \right) = 0.
\end{equation}  
\end{small}The following theorem elaborates the relationship between $\tilde {\mathbf{u}}_{r,g}^d\left( t \right)$ and the eigenvector samples $ {\tilde{\mathbf{u}}}_r^d\left( t \right)$ in EGVP-PRWCM.
\begin{theorem}\label{Theorem 2}
    The eigenvector ${{\tilde{\mathbf{u}}}_r^d\left( t \right)}$ obtained by \eqref{evd definition} and the eigenvector $\tilde {\mathbf{u}}_{r,g}^d\left( t \right)$ in  \eqref{eigenvector sample from channel relevance} are strictly correlated, i.e., $\tilde {\mathbf{u}}_{r,g}^d\left( t \right) = {\delta _r}\left( t \right){\tilde{\mathbf{u}}}_r^d\left( t \right)$, where ${{\delta }_r}\left( t \right)$ is the uncertainty factor caused by the EVD nature.
\end{theorem}
\begin{proof}
\quad \emph{Proof:} Please refer to Appendix \ref{Appendix Theorem 2}.
\end{proof}

Theorem \ref{Theorem 2} proves the equivalence of the eigenvector sampling between the two schemes. The complexity order of the EVD in EGVP-PRCWM reduce from ${{\cal O}}\left( {{N_f}^3{N_t}^3} \right)$ to ${{\cal O}}\left( {{N_r}^3} \right)$ compared to the EGVP-PRWCM scheme.

However, this scheme introduces an uncertainty factor ${{\delta }_r}\left( t \right)$. We prove that it can be eliminated by the phase alignment and normalization. Normally, the eigenvectors are orthogonal unit vectors:
\begin{small}
  \begin{equation}\label{channel weight unify}
\left\{ \begin{gathered}
  \left\langle {{\tilde{\mathbf{u}}}_r^d\left( t \right),{\tilde{\mathbf{u}}}_j^d\left( t \right)} \right\rangle  = 1,r = j \hfill \\
  \left\langle {{\tilde{\mathbf{u}}}_r^d\left( t \right),{\tilde{\mathbf{u}}}_j^d\left( t \right)} \right\rangle  = 0,r \ne j \hfill \\ 
\end{gathered}  \right.,\left\{ \begin{gathered}
  \left\langle {\tilde {\boldsymbol{\lambda }}_r^d\left( t \right),\tilde {\boldsymbol{\lambda }}_j^d\left( t \right)} \right\rangle  = 1,r = j \hfill \\
  \left\langle {\tilde {\boldsymbol{\lambda }}_r^d\left( t \right),\tilde {\boldsymbol{\lambda }}_j^d\left( t \right)} \right\rangle  = 0,r \ne j \hfill \\ 
\end{gathered}  \right..
\nonumber
\end{equation}   
\end{small}Therefore, the eigenvector in \eqref{eigenvector sample from channel relevance} needs to be normalized as $\overline {\mathbf{u}} _{r,g}^d\left( t \right) = {{\tilde {\mathbf{u}}_{r,g}^d\left( t \right)} \mathord{\left/
 {\vphantom {{\tilde {\mathbf{u}}_{r,g}^d\left( t \right)} {\left| {\tilde {\mathbf{u}}_{r,g}^d\left( t \right)} \right|}}} \right.
 \kern-\nulldelimiterspace} {\left| {\tilde {\mathbf{u}}_{r,g}^d\left( t \right)} \right|}}$ and phase-aligned according to the equation \eqref{eigenvector phase align}.

Then, the calibrated channel weight is
\begin{small}
  \begin{equation}\label{channel weight calibarated}
    \overline {\mathbf{\Lambda }} ^d\left( t \right) = \left[ {\begin{array}{*{20}{c}}
  {\overline {\mathbf{u}} _{1,g}^d\left( t \right)}&{\overline {\mathbf{u}} _{2,h}^d\left( t \right)}& \cdots &{\overline {\mathbf{u}} _{{N_r,g}}^d\left( t \right)} 
\end{array}} \right]{\tilde {\mathbf{H}}^d}{\left( t \right)^\dag }.
\end{equation}  
\end{small}
As a result, the uncertainty factor ${\delta _r}\left( t \right)$ becomes a random phase shift ${\overline \delta  _r}\left( t \right)$. 
%The system spectral efficiency (SE) is 
%\begin{small}
 %\begin{equation}\label{SE equation}
  %  {R_{{\text{se}}}}{\text{ = }}{\mathbb{E}}\left\{ {\sum\limits_{k = 1}^K {{\text{log}}\left( {1 + \frac{{\left\| {{{\mathbf{h}}_k}\left( t \right){\mathbf{f}}_{{\text{ezf}}}^k\left( t \right)} \right\|_2^2}}{{\sigma _k^2 + \sum\limits_{j \ne k}^K {{{\left| {{{\mathbf{h}}_k}\left( t \right){\mathbf{f}}_{{\text{ezf}}}^j\left( t \right)} \right|}^2}} }}} \right)} } \right\},
%\end{equation}   
%\end{small}
%where the precoder ${{\mathbf{f}}_{{\text{ezf}}}^k\left( t \right)}$ is usually chosen from ${\mathbf{u}}_1^d\left( t \right)$. 
Obviously, ${\overline \delta  _r}\left( t \right)$ does not affect the SE performance. However, the channel weight calculation \eqref{channel weight calibarated} relies on the DL channel estimation which may introduce a performance loss because of the channel estimation error. 

In the end, the eigenvectors are similarly predicted by \eqref{predicted eigenvector} and \eqref{channel weight interpolation}. The advantage of EGVP-PRCGM scheme over EGVP-PRWCM scheme lies in the reduced complexity of EVD when obtaining the channel weight samples. A more detailed complexity analysis and performance evaluation results of our proposed schemes are given in the next section.
\section{Performance analysis}\label{sec3}
In order to validate the advantages of our proposed precoding matrix prediction schemes, the complexity analysis is given and several numerical results are demonstrated.
\subsection{Complexity analysis}
The complexity of the precoding matrix prediction is analyzed within a period of time ${N_{{\text{svd}}}}{T_{{\text{svd}}}}$. The full-time SVD scheme updates the precoders in each subframe and the complexity is ${N_{{\text{svd}}}T_{{\text{svd}}}}{\cal{O}}\left( {{{\left( {{N_f}{N_t}} \right)}^3}} \right)$. However, the complexity of EGVP-PRWCM scheme is $\left( {{N_{{\text{svd}}}} + 2} \right)O\left( {{{\left( {{N_f}{N_t}} \right)}^3}} \right)$. The complexity of EGVP-PRCGM scheme is ${\mathcal{O}}\left( {{N_{{\text{svd}}}}{N_r}^5{{\left| {{{\mathcal{G}}_{i,j}}} \right|}^2_{{{\max }}}}{M^2} + 2{N_f}^3{N_t}^3} \right)$, where ${\left| {{{\mathcal{G}}_{i,j}}} \right|_{\max }} \leqslant {N_s}$ is the maximum ${{{\mathcal{G}}_{i,j}}}$ among all UEs.

Therefore, the complexity reduction from full-time SVD to EGVP-PRWCM is ${\mathcal{O}}\left( {{N_f}^3{N_t}^3} \right)\left( {{N_{{\text{svd}}}}{T_{{\text{svd}}}} - {N_{{\text{svd}}}} - 2} \right)$. It is always positive since $T_{{\text{svd}}},N_{{\text{svd}}} \ge 2$. Similarly, the complexity reduction from EGVP-PRWCM to EGVP-PRCGM is ${\mathcal{O}}\left( {{N_{{\text{svd}}}}\left( {{N_f}^3{N_t}^3 - {N_r}^5\left| {{{\mathcal{G}}_{i,j}}} \right|_{\max }^2{M^2}} \right)} \right)$, which is lower-bounded by ${\mathcal{O}}\left( {{N_{{\text{svd}}}}{N_f}^2{N_t}^2\left( {{N_f}{N_t} - {\kappa ^2}{N_r}^5} \right)} \right)$. In practical 5G systems, the configuration ${N_r} \leqslant 4,\kappa  \leqslant {1 \mathord{\left/
 {\vphantom {1 4}} \right.
 \kern-\nulldelimiterspace} 4},{N_t},{N_f} \geqslant 16$ is common. In this case, the EGVP-PRCGM scheme has a distinct complexity advantage over EGVP-PRWCM scheme.
\subsection{Numerical results and analysis}
Based on the cluster delay line-A (CDL-A) channel model \cite{3gpp901}, our proposed schemes are evaluated in several scenarios along with the benchmarks. The parameter configuration is shown in Table \ref{tab simulation config}. Both the BS and UEs are equipped with multiple antennas where $N_v$, $N_h$ are the number of antennas in the vertical direction and horizontal directions, respectively. $P_t$ is the number of polarization directions of the antennas.
%EVD cycle length $T_{\rm{svd}}= 5$ ms, CSI delay $T_d=5$ ms, matrix pencil prediction order $L_{{i,j}}=3$, eigenvector sample size $N_{\rm{svd}}=7$, feedback compression ratio $\kappa= {1 \mathord{\left/  {\vphantom {1 8}} \right.  \kern-\nulldelimiterspace} 8}$ are configured in 
The performances are measured in the system SE or the eigenvector prediction error (PE) metric like \cite{qin2023eigenvector}. 
\begin{table}[!t]
\centering \protect\protect\caption{Parameter Configurations in Simulations}
\label{tab simulation config}
\begin{tabular}{|c|c|}
%\hline Physical meaning & Default value\tabularnewline
\hline
%Channel model & CDL-A\tabularnewline
%\hline
Bandwidth & 20 $\rm{MHz}$\tabularnewline
\hline
UL/DL carrier frequency & 1.92 $\rm{GHz}$/2.11 $\rm{GHz}$\tabularnewline
\hline
Resource block & 51 RB\tabularnewline
\hline
Number of paths & 460\tabularnewline
\hline
\tabincell{c}{BS antenna \\configuration} & \tabincell{c}{$\left( {{N_v},{N_h},{P_t}} \right)= \left( {4,8,2} \right)$, \\ the polarization directions are ${0^\circ },{90^\circ }$}\tabularnewline
\hline
\tabincell{c}{UE antenna \\configuration} & \tabincell{c}{$\left( {{N_v},{N_h},{P_t}} \right)= \left( {1,2,2} \right)$, \\ the polarization directions are $\pm {45^\circ }$}\tabularnewline
\hline
Subframe duration & 1 $\rm{ms}$\tabularnewline
\hline
Number of UEs & 8\tabularnewline
\hline
Eigenvector sampling cycle & 5 $\rm{ms}$\tabularnewline
\hline
CSI delay & 5 $\rm{ms}$\tabularnewline
\hline
Prediction order & 3 \tabularnewline
\hline
Feedback compression ratio & ${1 \mathord{\left/
 {\vphantom {1 4}} \right.
 \kern-\nulldelimiterspace} 4}$ \tabularnewline
\hline
\end{tabular}
\end{table}
The performance upper bound is given when the DL CSI is perfectly known and the full-time SVD scheme is available. The rest benchmarks are all evaluated given the predicted CSI by the JADD method. The lazy SVD scheme is widely adopted in current 5G system \cite{3gpp214} and only updates the precoders every $T_{\rm{svd}}$ subframes. Wiener scheme utilizes the ${{L_w}}=2$ order Wiener-Hopf equation to predict the eigenvectors \cite{1992wierner}.
%\begin{equation}\label{wiener}
 %  {{{u}}_m}\left( {t + {T_d}} \right) = \sum\limits_{l = 1}^{{L_w}} {w\left( l \right){{{u}}_m}\left( {t - l} \right)},
%\end{equation}
%where the wiener coefficients ${w\left( l \right)}$ is calculated by history eigenvector samples \cite{1992wierner} . 
\begin{figure}[!t]
\centering
\includegraphics[width=2.8in]{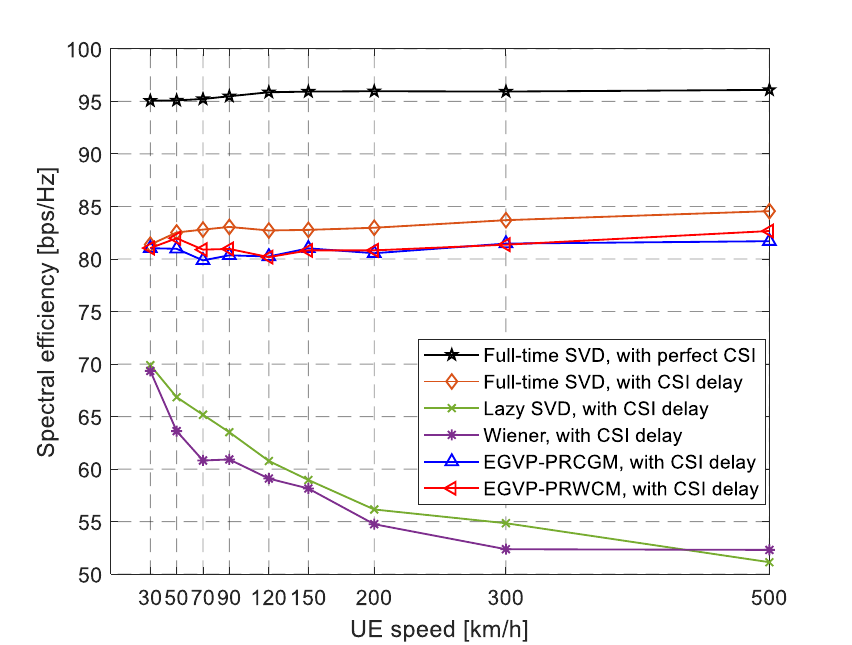}
\caption{SE performances under different UE speeds with noise-free channel samples, $\kappa={1 \mathord{\left/
 {\vphantom {1 4}} \right.
 \kern-\nulldelimiterspace} 4}$.}
\vspace{-0.3cm}
\label{fig_simulation_speed}
\end{figure}

First, the SE performances are evaluated under different UE speeds in Fig. \ref{fig_simulation_speed}. The performance gap between the upper bound and the other schemes is the result of the limited feedback compression ratio $\kappa$. Both EGVP-PRWCM and EGVP-PRCGM schemes closely approach the full-time SVD scheme with perfect CSI and outperform the other two benchmarks in low-speed and high-speed environments. The minor performance loss of EGVP-PRCGM compared to EGVP-PRCGM is caused by the additional channel weight calibration in \eqref{channel weight calibarated} and the DL channel estimation error.

\begin{figure}[!t]
\centering
\includegraphics[width=2.8in]{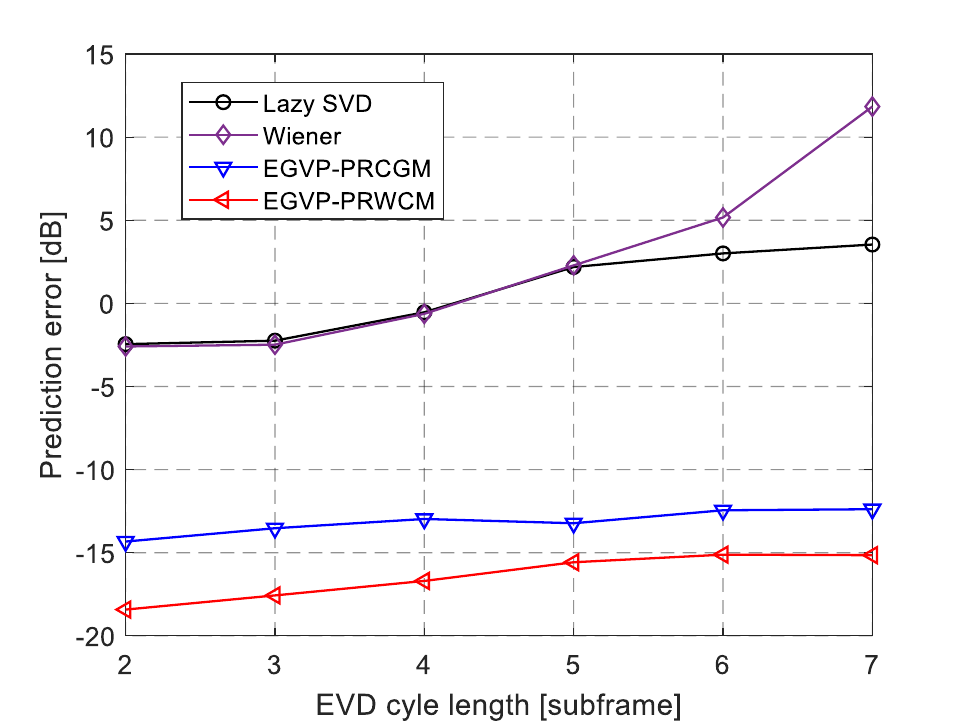}
\caption{Eigenvector PE performances under different EVD cycle lengths, $v=120$ km/h, $\kappa={1 \mathord{\left/
 {\vphantom {1 4}} \right.
 \kern-\nulldelimiterspace} 4}$.}
\vspace{-0.3cm}
\label{fig_simulation_cycle}
\end{figure}

Then, the PE performances under different EVD cycle lengths are shown in Fig. \ref{fig_simulation_cycle}.   The PE of our schemes increases slightly with the SVD cycle length. Therefore, our proposed schemes show better PE performances than the benchmarks.

\begin{figure}[!t]
\centering
\includegraphics[width=2.8in]{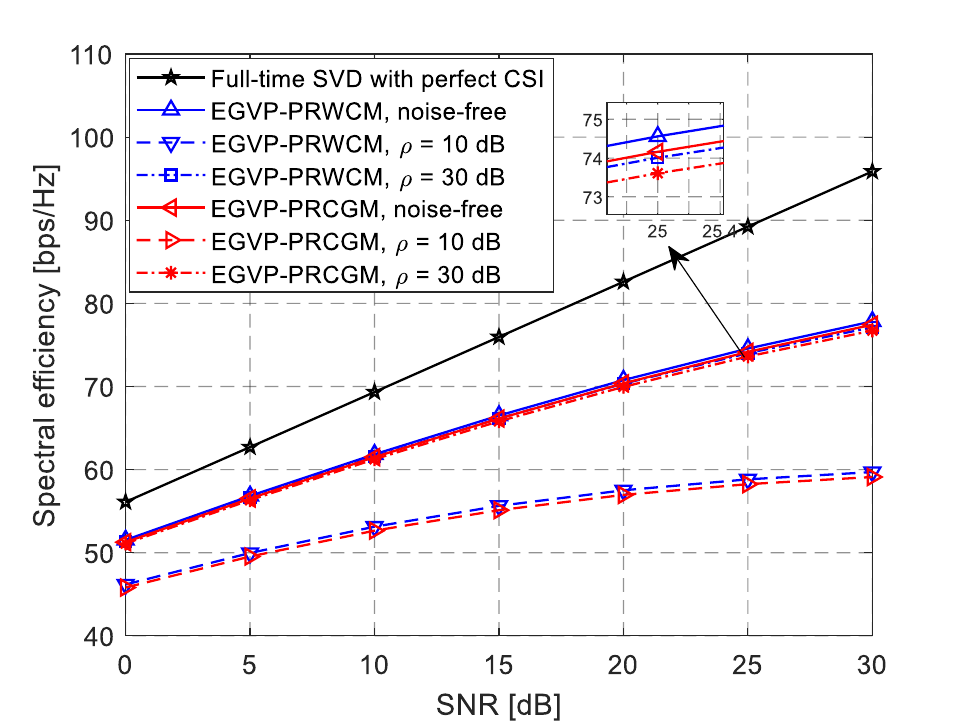}
\caption{SE performances under different channel sampling noises, $v=120$ km/h, $\kappa={1 \mathord{\left/
 {\vphantom {1 4}} \right.
 \kern-\nulldelimiterspace} 4}$.}
\vspace{-0.3cm}
\label{fig_simulation_noise}
\end{figure}

In the end, the SE performances under different channel sampling noises are demonstrated in Fig. \ref{fig_simulation_noise}. The channel sampling noise $\rho$ is measured by the power ratio of the channel estimation to the additive Gaussian noise. Although the channel estimation noise can be reduced by the denoising method in \cite{2009mdlNoise}, it has a significant impact on the SE performance. Fortunately, when $\rho$ increases to 30 dB, our scheme can achieve almost the same SE as in the noise-free case. 

\section{Conclusion}\label{sec4}
This paper focused on the precoding matrix prediction instead of channel estimation in FDD massive MIMO with mobility. Due to the limited computational capability of the BS in reality, the DL precoding matrix needs to be timely updated. Motivated by this, two partial reciprocity-based schemes were proposed. The general idea of the proposed schemes was based on our proved theorem that the DL precoders can be predicted through a closed-form eigenvector interpolation which was based on the periodic eigenvector samples. %Compared to the EGVP-PRWCM, EGVP-PRCGM further reduced the precoding complexity  by exploiting the partial reciprocity of the channel gram matrix. 
The complexity advantage of our schemes was proven and the performance improvement was evaluated with numerical results compared to the traditional full-time SVD scheme in various scenarios. 

% or


\appendices
\section{Proof of Theorem \ref{Theorem 1}}\label{Appendix Theorem 1}
\begin{proof}    
Based on the DL prediction \eqref{DL prediction}, the real DL channel is
\begin{small}
 \begin{equation}\label{dl channel with error}
    {\bf{h}}_r^d\left( t \right) = \sum\limits_{n = 1}^{{N_t}{N_f}} {g_{r,n}^d\left( t \right){{\bf{q}}_n}}  = \tilde {\bf{h}}_r^d\left( t \right) + \hat {\bf{h}}_r^d\left( t \right),
\end{equation}   
\end{small}where $g_{r,n}^d\left( t \right)$ is the amplitude of ${\bf{q}}_n$. The channel $\hat {\bf{h}}_r^d\left( t \right)$ represents the channel estimation error
\begin{small}
\begin{equation}\label{missing channel part}
    \hat {\mathbf{h}}_r^d\left( t \right) = \sum\limits_{n \notin {\mathcal{S}}_r^d} {\sum\limits_{m = 1}^M {a_{r,m}^d\left( n \right){e^{jw_{r,m}^d\left( n \right)t}}{{\mathbf{q}}_n}} } .
\end{equation}    
\end{small}Then the eigenvectors are 
\begin{small}
   \begin{equation}\label{eigenvector with error}
    {\bf{u}}_r^d\left( t \right) = \sum\limits_{j = 1}^{{N_r}} {\lambda _{r,j}^d\left( t \right)\left( {\tilde {\bf{h}}_r^d\left( t \right) + \hat {\bf{h}}_r^d\left( t \right)} \right)}  = \tilde {\bf{u}}_r^d\left( t \right) + \hat {\bf{u}}_r^d\left( t \right).
\end{equation} 
\end{small}

It was proved that the channel weight was asymptotically estimated by a complex exponential model \cite{qin2023eigenvector}
\begin{small}
 \begin{equation}\label{channel weight estimation model}
    \mathop {\lim }\limits_{{N_t},{N_f} \to \infty } {\mathbf{u}}_r^d\left( t \right) = \sum\limits_{j = 1}^{{N_r}} {\sum\limits_{l = 1}^{{L_{r,j}}} {b_{r,j}^d\left( l \right){e^{jw_{r,j}^d\left( l \right)t}}} \left( {\tilde {\mathbf{h}}_r^d\left( t \right) + \hat {\mathbf{h}}_r^d\left( t \right)} \right)}. 
\end{equation}   
\end{small}
Similarly, given the feedback compression ratio $\kappa$ in FDD, the estimated eigenvector can still be approximated by 
\begin{small}
  \begin{equation}\label{eigenvector estimation model}
    \tilde {\bf{u}}_r^d\left( t \right) = \sum\limits_{j = 1}^{{N_r}} {\lambda _{r,j}^d\left( t \right)\tilde {\bf{h}}_j^d\left( t \right)}  = \sum\limits_{j = 1}^{{N_r}} {\sum\limits_{l = 1}^{{L_{r,j}}} {b_{r,j}^d\left( l \right){e^{jw_{r,j}^d\left( l \right)t}}} \tilde {\bf{h}}_j^d\left( t \right)}.
\end{equation}  
\end{small}
Then, the PE upper bound is derived below. 
%The real eigenvector power is
%\begin{small}
 % \begin{equation}\label{eigenvector power total}
 %   \left\| {{\bf{u}}_r^d\left( t \right)} \right\|_2^2 \leqslant \sum\limits_{r = 1}^{{N_r}} {\left\| {\lambda _{r,j}^d\left( t \right)} \right\|_2^2\sum\limits_{n = 1}^{{N_f}{N_t}} {{{\left| {g_{r,n}^d\left( t \right)} \right|}^2}} } .
%\end{equation}  
%\end{small}
The estimated eigenvector satisfies
\begin{small}
 \begin{equation}\label{eigenvector power estimate}
   \left\| {\tilde {\bf{u}}_r^d\left( t \right)} \right\|_2^2 \leqslant \sum\limits_{r = 1}^{{N_r}} {\left\| {\lambda _{r,j}^d\left( t \right)} \right\|_2^2\sum\limits_{n \in {\mathcal{S}}_r^d} {{{\left| {g_{r,n}^d\left( t \right)} \right|}^2}} }.  
\end{equation}   
\end{small}Learned from Appendix C of \cite{2022ziao}, $M=1$ holds on when ${{N_t},{N_f} \to \infty }$. Considering the UL complex amplitude ${g_{r,n}^u\left( t \right)}$ has the following absolute value
\begin{small}
   \begin{equation}\label{ul complex amplitude}
    {\left| {g_{r,n}^u\left( t \right)} \right|^2} = a_r^u{\left( n \right)^*}{e^{ - jw_{r,m}^d\left( n \right)t}}a_r^u\left( n \right){e^{jw_{r,m}^d\left( n \right)t}} = {\left| {a_r^u\left( n \right)} \right|^2}.
\end{equation}     
\end{small}

Based on the channel partial reciprocity \cite{3gpp104}, the DL complex amplitude is
\begin{small}
 \begin{equation}\label{dl complex amplitude}
 {\left| {g_{r,n}^d\left( t \right)} \right|^2} = {\left| {a_{r}^d\left( n \right)} \right|^2} = {\left| {a_r^u\left( n \right)} \right|^2} .
\end{equation}   
\end{small}
According to UL the channel estimation power loss \eqref{angle-delay index selection}, the DL precoder PE is upper bounded by
\begin{small}
    \begin{align}
    {\mathbb{E}}\left\{ {\frac{{\left\| {{\mathbf{u}}_r^d\left( t \right) - \tilde {\mathbf{u}}_r^d\left( t \right)} \right\|_2^2}}{{\left\| {{\mathbf{u}}_r^d\left( t \right)} \right\|_2^2}}} \right\} &\le 
     {\mathbb{E}}\left\{ {\frac{{\sum\limits_{r = 1}^{{N_r}} {\left\| {\lambda _{r,j}^d\left( t \right)} \right\|_2^2\sum\limits_{n \notin {\mathcal{S}}_r^d} {{{\left| {g_{r,n}^d\left( t \right)} \right|}^2}} } }}{{\sum\limits_{r = 1}^{{N_r}} {\left\| {\lambda _{r,j}^d\left( t \right)} \right\|_2^2\sum\limits_{n = 1}^{{N_f}{N_t}} {{{\left| {g_{r,n}^d\left( t \right)} \right|}^2}} } }}} \right\} \nonumber\\&\le 1 - \eta 
\end{align}
\end{small}
Therefore, Theorem \ref{Theorem 1} is proved.
\end{proof}
\section{Proof of Theorem \ref{Theorem 2}}\label{Appendix Theorem 2}
\begin{proof}
According to the equation \eqref{evd definition}, the channel weight satisfies
\begin{small}
  \begin{equation}\label{channel weight definition}
\lambda _{r,j}^d\left( t \right){\chi _r}\left( t \right){\text{ }} = {\mathbf{h}}_j^d{\left( t \right)^H}\sum\limits_{i = 1}^{{N_r}} {\lambda _{r,i}^d\left( t \right){\mathbf{h}}_i^d\left( t \right)}  = \sum\limits_{i = 1}^{{N_r}} {s_{i,j}^d\left( t \right)\lambda _{r,i}^d\left( t \right)}.    
\end{equation}  
\end{small}
The above equations can be rewritten as
%\begin{small}
 \begin{equation}\label{channel weight definition ergodic}
\sum\limits_{i = 1}^{{N_r}} {{s_{i,1}}\left( t \right)\lambda _{r,i}^d\left( t \right)}  = {\chi _r}\left( t \right)\lambda _{r,j}^d\left( t \right),r,j \in \left\{ {1, \cdots {N_r}} \right\}.
\end{equation}   
%\end{small}
Comparing \eqref{channel weight definition ergodic} with \eqref{evd of channel gram matrix}, we can conclude that the wideband channel matrix  \eqref{evd definition} and the channel gram matrix \eqref{evd of channel gram matrix} share the same eigenvalues, i.e., ${\tilde \chi _r}\left( t \right) = {\chi _r}\left( t \right)$. Furthermore, ${\boldsymbol{\lambda }}_r^d\left( t \right)$ is also one of the eigenvectors of the channel gram matrix corresponding to the eigenvalue ${\tilde \chi _r}\left( t \right)$. Due to the EVD property, the eigenvectors associated with the same eigenvalue are strictly related, i.e., $\tilde {\boldsymbol{\lambda }}_r^d\left( t \right) = {\delta _r}\left( t \right){\boldsymbol{\lambda }}_r^d\left( t \right)$.  
Substitute ${\tilde {\boldsymbol{\lambda }}_r^d\left( t \right)}$ in \eqref{weight sample} and obtain
 \begin{equation}\label{weight diviation}
    \tilde {\mathbf{u}}_{r,g}^d\left( t \right) = {\delta _r}\left( t \right){\tilde{\mathbf{u}}}_r^d\left( t \right).
\end{equation}   
Therefore, Theorem \ref{Theorem 2} is proved.
\end{proof}

%\section*{Acknowledgment}

\ifCLASSOPTIONcaptionsoff
  \newpage
\fi



%\bibliographystyle{IEEEtran}
\bibliographystyle{IEEEtran}
\bibliography{IEEEabrv,reference}


%\end{IEEEbiography}


%\newpage



%\vfill


\end{document}